# Strong-Coupling Effects and Single-Particle Properties in an Ultracold Fermi Gas with Mass Imbalance

Ryo Hanai and Yoji Ohashi

*Department of Physics, Keio University, 3-14-1, Hiyoshi, Kohoku-ku, Yokohama 223-8522, Japan*

*E-mail: rhanai@rk.phys.keio.ac.jp*

We investigate single-particle properties of a strongly interacting ultracold Fermi gas with mass imbalance. Using an extended $T$-matrix theory, we calculate the density of states, as well as the single-particle spectral weight, in the unitarity limit above the superfluid phase transition temperature $T_c$. We show that the momentum regions where pairing fluctuations strongly affect single-particle excitations are different between light fermions and heavy fermions, reflecting the difference of the Pauli blocking effects between them. In addition, we obtain different pseudogap phenomena associated with pairing fluctuations in between the two components. Since the realization of a mass-imbalanced superfluid Fermi gas is an important challenge in this field, our results would contribute to the understanding of physical properties of the hetero-pairing state.
**KEYWORDS:** Fermi superfluid, mass imbalance, BCS-BEC crossover

## 1. Introduction

Since the realization of the BCS (Bardeen-Cooper-Schrieffer)-BEC (Bose-Einstein condensation) crossover in ultracold Fermi gases [1–3], the high controllability of this quantum system has attracted much attention [4]. More recently, JILA group has developed the photoemission-type experiment [5, 6], which enables us to study single-particle properties of this system in a microscopic manner. Indeed, using this technique, JILA has observed the pseudogap phenomenon in the crossover region [5, 6], where a gap-like structure appears in the single-particle excitation spectrum, even in the normal state. Stimulated by this experiment, several theoretical groups [7–12] have discussed the pseudogap physics in this system.

Besides the tunable interaction, "flexibility" with respect to the choice of particle species has also recently attracted much attention, as a route to explore further variety of Cooper pairings. Indeed, in addition to the "conventional" $^{40}$K-$^{40}$K [1] and $^6$Li-$^6$Li [2] Fermi gases, $^{40}$K-$^6$Li Fermi gases have been realized [13–15]. In this system, a Cooper pair naturally consists of different species (a $^{40}$K atom and a $^6$Li atom) having different masses, so that one can expect a Fermi superfluid, being different from the ordinary Fermi condensate of homo-pairs. Indeed, the so-called Sarma phase [16] has been predicted there [17]. In addition, since hetero-type Fermi superfluids have been discussed in various systems, such as an exciton gas [18], an exciton-polariton condensate [19] in a semiconductor, as well as color superconductivity in a dense quark matter [20], the study of the mass-imbalanced superfluid Fermi gas would also contribute to the further development of these fields. So far, although the formation of hetero-molecules has already been achieved [15], their superfluid instability is still in progress.

In this paper, we investigate a two-component unitary Fermi gas, to see how the mass imbalance affects strong-coupling properties of this system. In particular, we focus on the pseudogap phenomenon associated with strong pairing fluctuations. Using an extended $T$-matrix approximation [21, 22], which is valid for the case with mass imbalance [23], we calculate the density of states, as well as single-particle spectral weight, above the superfluid phase transition temperature $T_c$. We show that the Pauli blocking effect, as well as the pseudogap phenomenon, are very different between the light mass component and heavy mass component. Throughout this paper, we set $\hbar = k_B = 1$, and the system volume $V$ is taken to be unity, for simplicity.

## 2. Extended $T$-matrix theory for a mass-imbalanced Fermi gas

We consider a two-component Fermi gas with mass imbalance, described by the Hamiltonian,

$$H = \sum_{p,\sigma} \xi_{p\sigma} c^\dagger_{p\sigma} c_{p\sigma} - U \sum_{p,p',q} c^\dagger_{p+q/2,L} c^\dagger_{-p+q/2,H} c_{-p'+q/2,H} c_{p'+q/2,L}. \tag{1}$$

Here, $c^\dagger_{p,\sigma}$ is the creation operator of a Fermi atom, where the pseudospin $\sigma$ (= L, H) specifies the light (L) and heavy (H) mass component. The kinetic energy $\xi_{p\sigma} = p^2/(2m_\sigma) - \mu_\sigma$ is measured from the chemical potential $\mu_\sigma$ (where $m_\sigma$ is an atomic mass ($m_L \leq m_H$)). $-U(<0)$ is a tunable pairing interaction [3]. As usual, we measure the interaction strength in terms of the $s$-wave scattering length $a_s$, which is related to the pairing interaction $-U$ as $4\pi a_s/m = -U/[1 - U\sum_p m/p^2]$, where $m^{-1} = [m_L^{-1} + m_H^{-1}]/2$. The unitarity limit which we are considering in this paper is given by $a_s^{-1} = 0$. For simplicity, we deal with a uniform system, ignoring effects of a harmonic trap.

To include strong-coupling corrections to single-particle excitations in the presence of mass imbalance, we employ the extended $T$-matrix approximation developed in Refs. [21, 22]. In this theory, the self-energy correction $\Sigma_\sigma(p, i\omega_n)$ in the single-particle thermal Green's function $G^{-1}_{p\sigma}(i\omega_n) = i\omega_n - \xi_{p\sigma} - \Sigma_\sigma(p, i\omega_n)$ is given by [21, 22]

$$\Sigma_\sigma(p, i\omega_n) = T \sum_{q, i\nu_n} \Gamma(q, i\nu_n) G_{q-p, -\sigma}(i\nu_n - i\omega_n), \tag{2}$$

where $\omega_n$ and $\nu_n$ are fermion and boson Matsubara frequencies, respectively. In Eq. (2), $-\sigma$ means the opposite component to $\sigma$ (= L, H). The particle-particle scattering matrix $\Gamma(q, i\nu_n) = -U/[1 - U\Pi(q, i\nu_n)]$ describes pairing fluctuations, where $\Pi(q, i\nu_n) = T \sum_{p, i\omega_n} G^0_{p+\frac{q}{2}, L}(i\nu_n + i\omega_n) G^0_{-p+\frac{q}{2}, H}(-i\omega_n)$ is a pair-propagator (where $G^0_{p\sigma}(i\omega_n) = [i\omega_n - \xi_{p\sigma}]^{-1}$ is a Green's function of a free $\sigma$-spin atom).

As usual, the superfluid transition temperature $T_c$ is determined from the Thouless criterion [24], $\Gamma(q = 0, i\nu_n = 0)^{-1} = 0$, together with the number equations $N_\sigma = T \sum_{p, i\omega_n} G_{p\sigma}(i\omega_n)$ under the unpolarized condition $N_L = N_H \equiv N/2$, to determine $(T_c, \mu_L, \mu_H)$ self-consistently. Above $T_c$, we only solve the number equations for $N_\sigma$, to determine $\mu_\sigma$. The single-particle spectral weight $A_\sigma(p, \omega)$, as well as the density of states $\rho_\sigma(\omega)$, are evaluated from the analytic continued Green's function as, respectively,

$$A_\sigma(p, \omega) = -\frac{1}{\pi} \text{Im}[G_{p\sigma}(i\omega_n \to \omega + i\delta)], \tag{3}$$

$$\rho_\sigma(\omega) = \sum_p A_\sigma(p, \omega). \tag{4}$$

## 3. Single-particle properties of a mass-imbalanced unitary Fermi gas

Figure 1 shows the single-particle spectral weight $A_\sigma(p, \omega)$ in the normal state of a unitary Fermi gas, when $m_L/m_H = 0.2$. In the light mass component shown in panel (a), one sees that the peak width rapidly changes around the Fermi momentum $k_F \equiv (3\pi^2 N)^{1/3}$, namely, the peak is sharp (broad) when $p \lesssim k_F$ ($p \gtrsim k_F$). In contrast, such a remarkable change of the peak width is not seen around $p = k_F$ in the heavy mass component (panel (b)).

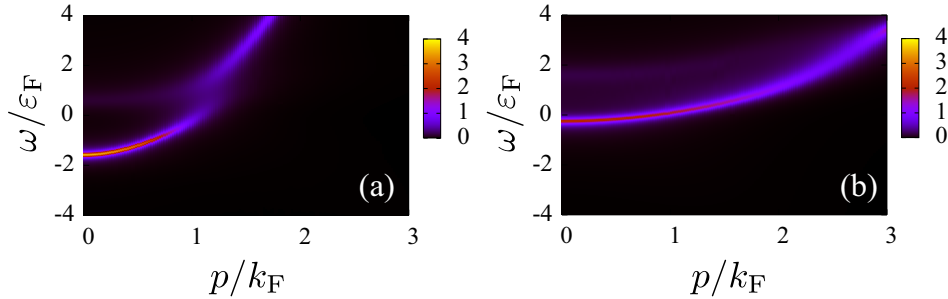

**Fig. 1.** Calculated intensity of the single-particle spectral weight $A_\sigma(p,\omega)$ in the unitarity limit ($(k_F a_s)^{-1} = 0$). (a) Light mass component. (b) Heavy mass component. We set $m_L/m_H = 0.2$, and $T = 0.21 T_F$, where $T_F = (3\pi^2 N/2\sqrt{2})^{2/3}/m)$. The spectral intensity is normalized by the inverse of the Fermi energy $\varepsilon_F$ ($= T_F$). The momentum is normalized by the Fermi momentum $k_F$.

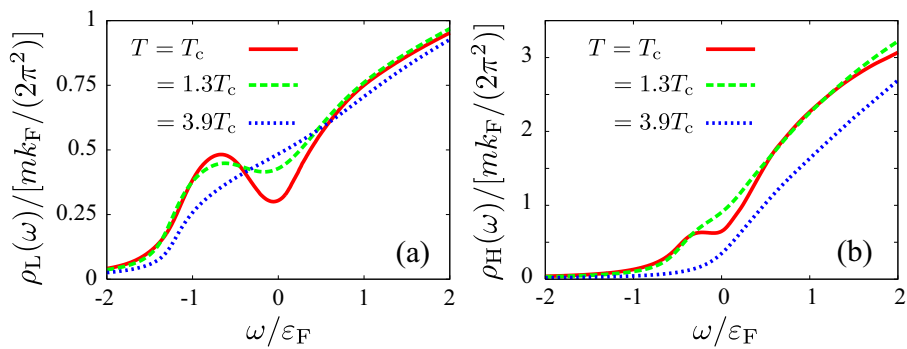

**Fig. 2.** Calculated density of states in the unitarity limit. (a) Light mass component. (b) Heavy mass component. We take $m_L/m_H = 0.4$. In this case, the superfluid phase transition temperature equals $T_c = 0.15 T_F$.

To understand the reason for this difference, it is convenient to consider the non-interacting case. In this simplest case, while both components have a common Fermi momentum $k_F$ at $T = 0$, the Fermi temperature of the heavy mass component $T_F^H = k_F^2/(2m_H)$ is lower than that of the light mass component $T_F^L = k_F^2/(2m_L)$, reflecting $m_H > m_L$. Then, noting that the scaled temperature $T/T_F^\sigma$ determines the Fermi degeneracy of the $\sigma$-component, one can expect that quantum effects are more remarkable in the light mass component than in the heavy one, for a given temperature $T$. That is, in the light mass component, particle scatterings below the Fermi level are suppressed by the Pauli blocking effect, leading to the sharp spectral peak in this regime. Since the Pauli blocking does not work above the Fermi level, light fermions in this regime have broad spectral widths, as shown in Fig.1(a). On the other hand, since the heavy mass component is effectively closer to the classical regime ($T \sim T_F^H$), the Fermi distribution function is more broadened by thermal effects than the case of light fermions. Thus, the Pauli blocking effect below the Fermi level is weaker than the case of the light mass component, so that the spectral peak width does not rapidly change around the Fermi level, as shown in Fig.1(b).

The importance of the scaled temperature $T/T_F^\sigma$ is also seen in the pseudogap phenomenon, as shown in Fig.2. With increasing temperature from $T_c$, effects of thermal fluctuations become stronger in the heavy mass component, due to the higher scaled temperature $T/T_F^H$ ($> T/T_F^L$). Thus, the pseudogap in the density of states $\rho_H(\omega)$ is soon smeared out to disappear above $T_c$, as shown in Fig.2(b). In contrast, panel (a) shows that the pseudogap in the light mass component remains, even at $T = 1.3T_c$ (where the pseudogap in the heavy component has disappeared).

## 4. Summary

To summarize, we have discussed a mass imbalanced Fermi gas in the unitarity limit. Using an extended $T$-matrix theory, we clarified how the mass imbalance affects single-particle properties of this system in the normal state. Since heavy mass component effectively feels a higher *scaled* temperature $T/T_F^H$ ($> T/T_F^L$) than the light mass component, the former is more sensitive to thermal effects. This naturally leads to different momentum dependences of the peak width in the spectral weight $A_\sigma(p, \omega)$ around the Fermi level. In addition, we also showed that the pseudogap structure in the density of states disappears sooner in the heavy mass component than in the light mass component, as one increases the temperature from $T_c$. Since strongly interacting mass-imbalanced Fermi superfluids have recently attracted much attention in various fields, such as a $^{40}$K-$^6$Li gas mixture, an exciton (polariton) gas in a semiconductor, as well as color superconductivity in a dense quark matter, our results would contribute to the development of these novel Fermi superfluids.

## Acknowledgements

We thank H.Tajima for useful discussions. R.H. was supported by KLL PhD. Program Research Grant. Y.O. was supported by Grant-in-Aid for Scientific research from MEXT in Japan (25400418, 25105511, 23500056).

## References

[1] C. A. Regal, M. Greiner, and D. S. Jin, Phys. Rev. Lett. **92**, 040403 (2004).
[2] M. Zwierlein, C. Schunck, C. Stan, Phys. Rev. Lett. **94**, 180401 (2005).
[3] C. Chin, P. Julienne, E. Tiesinga, Rev. Mod. Phys. **82**, 1225 (2010).
[4] For a review, see, S. Giorgini, L. Pitaevskii, S. Stringari, Rev. Mod. Phys. **80**, 1215 (2008).
[5] J. Stewart, J. Gaebler, and D. Jin, Nature **454**, 744 (2008).
[6] J. Gaebler, J. Stewart, T. Drake, D. Jin, A. Perali, P. Pieri, and G. Strinati, Nature Phys. **6**, 569 (2010).
[7] S. Tsuchiya, R. Watanabe, Y. Ohashi, Phys. Rev. A **80**, 033613 (2009); **84**, 043647 (2011).
[8] Q. Chen and K. Levin, Phys. Rev. Lett. **102**, 190402 (2009).
[9] H. Hu, X.-J. Liu, P. D. Drummond, and H. Dong, Phys. Rev. Lett. **104**, 240407 (2010).
[10] E. Mueller, Phys. Rev. A **83**, 053623 (2011).
[11] A. Perali, F. Palestini, P. Pieri, G. Strinati, J. Stewart, J. Gaebler, T. Drake, and D. Jin, Phys. Rev. Lett. **106**, 060402 (2011).
[12] R. Watanabe, S. Tsuchiya, Y. Ohashi, Phys. Rev. A **86**, 063603 (2012).
[13] E. Wille, F. Spiegelhalder, G. Kerner, D. Naik, A. Trenkwalder, G. Hendl, F. Schreck, R. Grimm, T. Tiecke, J. Walraven, S. Kokkelmans, E. Tiesinga, P. S. Julienne, Phys. Rev. Lett. **100**, 053201 (2008).
[14] M. Taglieber, A. Voigt, T. Aoki, Phys. Rev. Lett. **100**, 010401 (2008).
[15] A. C. Voigt, M. Taglieber, L. Costa, T. Aoki, Phys. Rev. Lett. **102**, 020405 (2009).
[16] G. Sarma, J. Phys. Chem. Solids **24**, 1029 (1963).
[17] J. Baarsma, K. Gubbels, H. Stoof, Phys. Rev. A **82**, 013624 (2010).
[18] K. Yoshioka, E. Chae, M. Kuwata-Gonokami, Nature Comm. **2**, 328 (2011).
[19] J. Kasprzak, M. Richard, S. Kundermann, A. Baas, P. Jeambrun, J. Keeling, F. Marchetti, M. Szymańska, R. Andre, J. Staehli, V. Savona, P. B. Littlewood, B. Deveaud, L. Dang, Nature **443**, 409 (2006).
[20] B. Barrois, Nucl. Phys. B **129**, 390 (1977).
[21] T. Kashimura, R. Watanabe, Y. Ohashi, Phys. Rev. A **86**, 043622 (2012).
[22] R. Hanai, T. Kashimura, R. Watanabe, D. Inotani, Y. Ohashi, J. Low Temp. Phys. **171**, 389 (2013).
[23] The Gaussian fluctuation theory, as well as the ordinary (non-self-consistent) $T$-matrix theory, that are known to be valid for the mass-balanced case, unphysically give double-valued $T_c$ in the crossover region. For more details, we refer to Ref. [22].
[24] D. J. Thouless, Ann. Phys. (N.Y.) **10**, 553 (1960).